\documentclass[%
reprint,
superscriptaddress,
 amsmath,amssymb,
 aps,
]{revtex4-1}

\usepackage{graphicx}
\usepackage{dcolumn}
\usepackage{bm}
\usepackage[colorlinks, linkcolor=red, anchorcolor=blue,urlcolor=blue, citecolor=blue]{hyperref}
\usepackage{color}

\newcommand{\beq}{\begin{equation}}
\newcommand{\eeq}{\end{equation}}
\newcommand{\beqa}{\begin{eqnarray}}
\newcommand{\eeqa}{\end{eqnarray}}

\newcommand{\arccot}{\mathrm{arccot}\,}

\makeatletter
\newcommand{\Rmnum}[1]{\expandafter\@slowromancap\romannumeral #1@}
\makeatother

\begin{document}

\title{Fast and robust control of two interacting spins}

\author{Xiao-Tong Yu}
\affiliation{Department of Physics, Shanghai University, 200444 Shanghai, People's Republic of China}

\author{Qi Zhang}
\affiliation{Department of Physics, Shanghai University, 200444 Shanghai, People's Republic of China}

\author{Yue Ban}
\affiliation{Department of Electronic Information Materials, Shanghai University, 200444 Shanghai, People's Republic of China}
\affiliation{Instituto de Ciencia de Materiales de Madrid, CSIC, Sor Juana In\'{e}s de la Cruz 3, E-28049 Madrid, Spain}

\author{Xi Chen}
\email{xchen@shu.edu.cn}
\affiliation{Department of Physics, Shanghai University, 200444 Shanghai, People's Republic of China}

\date{\today}

\begin{abstract}
Rapid preparation, manipulation, and correction of spin states with high fidelity are requisite for quantum information processing and quantum computing.
In this paper, we propose a fast and robust approach for controlling two spins with Heisenberg and Ising interactions.
By using the concept of shortcuts to adiabaticity, we first inverse design the driving magnetic fields for achieving fast spin flip or generating the entangled Bell state, and further optimize them with respect to the error and fluctuation.
In particular, the designed shortcut protocols can efficiently suppress the unwanted transition or control error induced by anisotropic antisymmetric Dzyaloshinskii-Moriya exchange. Several examples and comparisons are illustrated, showing the advantages of our methods. Finally, we emphasize that the results can be naturally extended to multiple interacting spins and other quantum systems in an analogous fashion.
\end{abstract}


\maketitle

\section{Introduction}


Efficient initialization and manipulation of quantum states have been long pursued in the fields of quantum optics, quantum control, and even quantum simulation,
due to their significance for information storage, processing and computing in various systems \cite{Bookquantum,BookLoss}. Along this research line,
different approaches including $\pi$ (or $\pi/2$) resonant pulse \cite{Eberly}, adiabatic passages and its variations \cite{Bergmann-Rev1,Bergmann-Rev2}
have been proposed to achieve such desirable goal. As compared to simple resonant pulses, adiabatic passages are usually robust against the systematic errors,
but take long time, due to adiabatic criteria. The shortcoming is that the long-time state evolution will be spoiled due to decoherence effects in a noisy environment.
To remedy it, several complementary methods, i.e., composite pulses \cite{NMR,Torosov-PRL,Genov} and optimal control \cite{Tannor,Boscain,review},
have been proposed.

In recent years, an alternative concept of ``shortcuts to adiabaticity" (STA) \cite{RevSTA} has been put forward
to speed up the slow adiabatic process without final excitation, with the broad applications, ranging from atomic, molecular, optical physics to solid-state physics. Among them, two specific protocols, inverse engineering \cite{PRL104} and counter-diabatic driving
(or equivalently quantum transitionless algorithm)\cite{Rice,Berry09,PRL105}, are popular for different motivations and proposals. In principle these two methods
are mathematically equivalent, but the physical implementations are quite different \cite{XEMuga}. For instance, ultrafast internal state manipulation \cite{Oliver,Suter,Li,NV}
and ion transport in phase space \cite{Kim} have been realized in the state-of-the-art experimental implementations.
The disadvantage, however, is that the counter-diabatic terms are sometimes infeasible or completely unphysical \cite{SaraPRL}.
Besides, the inverse engineering stems from Lewis-Riesenfeld (LR) dynamics invariant \cite{LR}, which was first proposed for fast frictionless atom cooling in harmonic traps \cite{PRL104}
and demonstrated experimentally as well \cite{Nice1,Nice2}. More importantly, such shortcut provides more freedom for further optimization
when the dynamics is designed only from the appropriate initial and final boundary conditions \cite{2012njp,PRALu,PRLDijon}.
Actually, other methods can also reproduce the similar results from the inverse engineering strategy \cite{PRLAo,SarmaPRL,Vitanov17,JieSong}.

Particularly, the applications of STA in spin-1/2 system are always amazing, since
the (optimal) control of its dynamics is demanding in nuclear magnetic resonance (NMR) \cite{Assemat,Sugny,Boozer}, nitrogen-vacancy (NV) centre in diamond \cite{Suter,NV} and quantum dots \cite{YuePRL,YueSP,WangXin},
and such system also resembles ubiquitous two-level quantum systems, for instance, superconducting circuits \cite{Barnes}, and optomechanical systems \cite{ClerkNC}.
For single spin, the controllable magnetic fields are shaped, respectively, from STA methods of counter-diabatic driving \cite{Berry09}, inverse engineering \cite{PRLAo,Tokatly,QiSci,Santos} and
fast-forward scaling \cite{Masuda,Takahashi-1,Takahashi-2}. Besides, the manipulation of two spins with isotropic and anisotropic exchange interactions become naturally attractive for entangled state generation and quantum annealing.
The system of two interacting spins with four internal levels, satisfying SU(4) Lie algebra, allows one to design the STA by using LR invariant \cite{YiDun}.
Recently, the inverse engineering for four-level systems with specific coupling configurations is further developed by using four dimensional double rotation \cite{YCLi}.
But, in certain cases the four-level system can be simplified to three-level or two-level systems in terms of adiabatic elimination.
As a consequence, the counter-diabatic terms \cite{Masuda17,Shi17} are easily calculated and implemented. Again, we shall emphasize that the inverse engineering \cite{QiSci,Sarmaentangle} really provides an efficient way for speeding up the conventional adiabatic passage \cite{bell01}, and producing the efficient quantum gates
as well \cite{Santos}.
However, the optimization of STA with respect to the error and fluctuation, contributed from control field and perturbative interaction, has not been explored yet,
which is an important issue in practice.

\begin{figure}[t]
\begin{center}
\scalebox{0.6}[0.6]{\includegraphics{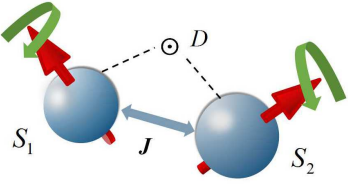}}
\caption{Schematic diagram for two-spin system with the Heisenberg or Ising interactions $J$ and Dzyaloshinskii-Moriya interaction $D$, anisotropic antisymmetric exchange.}\label{fig.spins}
\end{center}
\end{figure}

In this article, we shall study systematically the optimal control of two interacting spins through time-dependent magnetic fields by using the technique of STA.
For simplicity, we consider fast control of spin states in two coupled spins systems, see Fig. \ref{fig.spins}, in presence of isotropic Heisenberg or anisotropic
Ising interaction, and the controllable magnetic fields are designed inversely correspondingly. By combining with time-dependent perturbation theory,
the spin dynamics is further optimized with respect to the errors and fluctuations. Moreover,
we shall consider the influence of Dzyaloshinskii-Moriya (DM) interaction in two-spin systems, which is an anisotropic antisymmetric interaction,
due to spin-orbit coupling. This non-negligible term contributes to control error in quantum information processing \cite{Guerrero}.
Our strategy is to treat it as perturbative error, rather than counter-diabatic term \cite{Shi17}, and to improve the fidelity
by suppressing the unwanted transition or cancelling the control error.
All these results are demonstrated by numerical examples, and compared with the resonant and composite pulses,
showing the advantage of robustness.

The paper is organized as follows. In Sec. \ref{Heisenberg}, we consider the two Heisenberg-interacting spins to design
the fast and robust spin flip with systematic errors and perturbative DM interactions. In Sec. \ref{Ising}
we design such shortcuts again for two Ising-interacting spins in presence of systematic errors and DM interaction,
to generate the entangled Bell state with a short time scale.
We finally present the discussion and summary in Sec. \ref{discussion} and \ref{conclusion}.

\section{Two Heisenberg-interacting Spins}
\label{Heisenberg}
First of all, we consider the two-spin system of spin quantum numbers $\vec{S}_1$ and $\vec{S}_2$ with isotropic exchange coupling in presence of a time-dependent magnetic field, which is described by the Hamiltonian
\begin{equation}
\label{H1}
H (t)=J \vec{S}_1\cdot\vec{S}_2+\vec{B}(t)\cdot (\vec{S}_1+\vec{S}_2), 
\end{equation}
where $J>0$  describes antiferromagnetic coupling ($J <0$ ferromagnetic coupling), and $\vec{B} (t)$ is rotating magnetic field with
three components, $B_{i}$ ($i=x,y,z$). (Noting that we consider the isotropic one for simplicity, and
there also exist other anisotropic Heisenberg model, see Appendix \ref{appendix}.)
In the basis of $\{S_z, S^2 \}$,
\begin{eqnarray}
\label{basis}
  &|&\psi_{1,1}\rangle = |\uparrow\uparrow\rangle, \\
  &|&\psi_{1,0}\rangle = \frac{1}{\sqrt2}(|\uparrow\downarrow\rangle+|\downarrow\uparrow\rangle), \\
  &|&\psi_{0,0}\rangle = \frac{1}{\sqrt2}(|\uparrow\downarrow\rangle-|\downarrow\uparrow\rangle), \\
  \label{basis4}
  &|&\psi_{1,-1}\rangle = |\downarrow\downarrow\rangle,
\end{eqnarray}
the Hamiltonian has the following matrix form (by setting $\hbar\equiv1$):
\begin{equation}
  H(t)=\left(
         \begin{array}{cccc}
           \frac{J}{4}+ B_z  & \frac{B_x - i B_y}{\sqrt{2}} & 0 & 0 \\
           \frac{B_x + i B_y}{\sqrt{2}} &  \frac{J}{4} & 0 & \frac{B_x - i B_y }{\sqrt{2}} \\
           0 & 0 &  -\frac{3 J}{4} & 0 \\
           0 & \frac{B_x + i B_y}{\sqrt{2}} & 0 &  \frac{J}{4} - B_z \\
         \end{array}
       \right),
\end{equation}
where the time dependence is omitted for simplicity.
By shifting the energy $J/4$, we can further simplify the Hamiltonian as follows,
\begin{equation}
\label{matrixH1}
  H(t)=\left(
         \begin{array}{cccc}
           B_z  & \frac{B_x - i B_y}{\sqrt{2}} & 0 & 0 \\
           \frac{B_x + i B_y}{\sqrt{2}} &  0 & 0 & \frac{B_x - i B_y }{\sqrt{2}} \\
           0 & 0 &  -J & 0 \\
           0 & \frac{B_x + i B_y}{\sqrt{2}} & 0 &  - B_z \\
         \end{array}
       \right).
\end{equation}
Since there exists one level decoupled to the other three,
the Hamiltonian can be further reduced to
\begin{equation}
\label{Ha}
  H (t)=\left(
      \begin{array}{ccc}
        \Delta & \frac{1}{\sqrt{2}}\Omega & 0 \\
        \frac{1}{\sqrt{2}}\Omega  & 0 & \frac{1}{\sqrt{2}}\Omega \\
        0 & \frac{1}{\sqrt{2}}\Omega & -\Delta \\
      \end{array}
    \right),
\end{equation}
when we impose $\Omega =B_x $, $\Delta=B_z$, $B_y =0$.
Here $\Omega$ and $\Delta$ refer to two components of magnetic fields, which resemble the Rabi freqeuncy and detuning in quantum optics as well.
With this interaction in such type of two coupled spins, one cannot reach the Bell state $|\psi_{1,0}\rangle$,
from $| \psi_{1,1}\rangle$, since there no energy gap between these two states. Instead,
the time-dependent magnetic field can drive the states from $| \psi_{1,1}\rangle$ to $| \psi_{1,-1}\rangle$.
But the adiabatic passage takes long time to satisfy the widely used adiabatic condition, $T^{ad} \gg \Omega_0/ a$ \cite{Du}.
As an example, Landau-Zener scheme, $\Omega= \Omega_0$ and $\Delta= a (t-T/2)$, requires $T^{ad} \simeq 20$,
with constant Rabi freqency $\Omega_0 =8$ and chirp $a =4$. Our motivation is to speed up and optimize
the spin-flip process by using inverse engineering.

This Hamiltonian (\ref{Ha}) for three-level model
maps into spin-1 system, and can thus be rewritten as
$ H = \Omega J_x + \Delta J_z$,
where $J_{\nu}$ ($\nu = x,y,z$) are spin-1 generator matrices, satisfying SU(2) algebra,
$[J_\mu,J_\nu]=i J_\gamma \varepsilon_{\mu \nu \gamma}$, with
the structure constants $\varepsilon_{\mu \nu \gamma}$ (see Ref. \cite{Lie}).
Consequently, the dynamic invariant $I(t)$, satisfying $d I(t)/dt \equiv \partial I(t)/ \partial t + (1/ i \hbar) [H(t),I(t)] =0 $,
can be constructed as \cite{LR}
\begin{equation}
\label{dotI1}
  I(t)= B_0 \left( \begin{array}{ccc}
                          \cos\theta & \frac{1}{\sqrt2}\sin\theta e^{-i\beta} & 0 \\
                     \frac{1}{\sqrt2}\sin\theta e^{i\beta} & 0 & \frac{1}{\sqrt2}\sin\theta e^{-i\beta} \\
                          0 & \frac{1}{\sqrt2}\sin\theta e^{i\beta} & -\cos\theta \\
                        \end{array}
                      \right),
\end{equation}
where $B_0$ is a constant magnitude of magnetic field, guaranteeing the same dimension as $H(t)$.
The eigenstates of the invariant $I(t)$, $I(t) | \phi_n(t) \rangle = \lambda_n
|\phi_n (t) \rangle$ (we use the labels $n=0,1,2$), can be easily obtained as
\begin{eqnarray}
\label{phi3-1}
  |\phi_{0}(t)\rangle &=\frac{1}{\sqrt2}& \left(
                                            \begin{array}{c}
                                              -\sin{\theta}e^{-i\beta} \\
                                              \sqrt{2}\cos{\theta} \\
                                              \sin{\theta}e^{i\beta} \\
                                            \end{array}
\right),\\
\label{phi3-2}
  |\phi_{1}(t)\rangle &=& \left(
                        \begin{array}{c}
                          \cos^2{\frac{\theta}{2}}e^{-i\beta} \\
                          \frac{1}{\sqrt{2}}\sin\theta \\
                          \sin^2{\frac{\theta}{2}}e^{i\beta} \\
                        \end{array}
                      \right),\\
\label{phi3-3}
  |\phi_{2}(t)\rangle &=& \left(
                        \begin{array}{c}
                          \sin^2{\frac{\theta}{2}}e^{-i\beta} \\
                          -\frac{1}{\sqrt{2}}\sin\theta \\
                          \cos^2{\frac{\theta}{2}}e^{i\beta} \\
                        \end{array}
                      \right),
\end{eqnarray}
with corresponding eigenvalues $\lambda_0=0$ and $\lambda_{1,2} = \pm B_0/2$.
Based on LR theory, the dynamics of such three-level system,
described by the time-dependent Schr\"{o}dinger equation, $i \hbar \partial_t
| \Psi(t) \rangle = H | \Psi(t) \rangle$, is in general governed by
the superposition of orthogonal ``dynamical mode" \cite{LR}
\begin{equation}
\label{solution}
|\Psi (t)\rangle= \Sigma_n c_n e^{i\gamma_n}|\phi_{n}(t)\rangle,
\end{equation}
where $c_n$ is time-independent constant and the LR phases $\gamma_n$ are solved as $\gamma_0=0$,
\begin{equation}
\label{phase}
\gamma_{1,2} = \pm \int_0^t dt' \left(\frac{\dot\theta \cot\beta}{\sin\theta}\right).
\end{equation}
From the condition for dynamical invariant, we have following auxiliary differential equations:
\begin{eqnarray}
\label{dotflip-1}
\dot\theta  &=& -\Omega \sin\beta, \\
\label{dotflip-2}
\dot\beta  &=& \Delta  - \Omega \cot\theta  \cos\beta,
\end{eqnarray}
by which the magnetic field components, $\Delta$ and $\Omega$, are connected with two parameters $\theta$ and $\beta$.
Now it is ready to apply the inverse engineering for achieving the fast spin flip from $|\psi_{1,1}\rangle$ to $|\psi_{1,-1}\rangle$ within short time.
The perturbation theory along the single dynamical mode $| \phi_1 (t) \rangle$,
and impose the boundary conditions,
\beqa
\label{bc1}
\theta(0)=0, ~~ \theta(T)=\pi
\\
\label{bc2}
\dot\theta(0)=0,~~ \dot\theta(T)=0.
\eeqa
Note that the boundary conditions (\ref{bc1}) are necessary for spin flip, while the others (\ref{bc2}) make the fields smooth at the edges. Once $\theta$ and $\beta$ are interpolated by polynomial ansatz \cite{Sarmaentangle} with the appropriate boundary conditions (\ref{bc1}) and (\ref{bc2}), the magnetic fields can be inversely designed from Eqs. (\ref{dotflip-1}) and (\ref{dotflip-2}).
In principle, the inverse engineering has more flexibilities than counter-diabatic driving, since there are thousands of possible paths
to connect the boundary conditions. One has to optimize the shortcut by combining with the time-dependent perturbation theory \cite{2012njp}, or numerical recipe \cite{NJPSherson}.
In what follows, we shall optimize the magnetic fields with respect to the
systematic error, and also suppress the unwanted transition, induced by anisotropic antisymmetric DM interaction.

\subsection{Systematic Error}
The existence of noises, errors, and fluctuations is unavoidable during the state control for most practical quantum systems.
The optimization of STA with respect to systematic errors is helpful, since the rotating magnetic field might have imperfection,
resulting in the shift $\delta$ in the amplitude of $\vec{B}$. Here we describe such sole systematic error by the perturbative Hamiltonian,
that is,
\begin{equation}
 H'(t)=\left(
      \begin{array}{ccc}
        \delta\Delta & \frac{1}{\sqrt{2}} \delta\Omega & 0 \\
        \frac{1}{\sqrt{2}}\delta\Omega  & 0 & \frac{1}{\sqrt{2}}\delta\Omega \\
        0 & \frac{1}{\sqrt{2}} \delta\Omega & - \delta\Delta \\
      \end{array}
    \right),
\end{equation}
which means the two components of magnetic field are simultaneously shifted
as $\Delta \rightarrow \Delta(1+\delta)$ and $\Omega \rightarrow \Omega(1+\delta)$.

By using time-dependent perturbation theory, we have
\begin{equation}
\label{state}
\begin{split}
  &|\Psi(T)\rangle=|\Psi_1(T)\rangle-i \int_0^T dt U_0(T,t) H'|\Psi_1(t)\rangle\\
  &-\int_0^T dt\int_0^tdt'U_0(T,t)H'U_0(t,t')H'|\Psi_1(t')\rangle+...,
  \end{split}
\end{equation}
where the unperturbed time evolution operator $U_0(T,t)= \sum_n |\Psi_n (T)\rangle \langle \Psi_n (t) |$ with
$|\Psi_n (t) \rangle =e^{i \gamma_n (t)} |\phi_n (t) \rangle$
where $n=0,1,2$.

The fidelity to find the final state $|\psi_{1,-1}\rangle$ from initial state $|\psi_{1,1}\rangle$ along one of the dynamical modes, $|\Psi_1 (t) \rangle $, is defined as
$F=|\langle \Psi_1 (T)| \Psi(T) \rangle|^2 $, and can be further estimated as, by keeping the second order,
\beq
\label{p1}
F \simeq 1- \sum_{n  \neq 1} \left|\int_0^T dt\langle\Psi_1 (t)| H' |\Psi_n (t)\rangle\right|^2.
\eeq
By defining the systematic error sensitivity as \cite{2012njp}
\begin{equation}
\label{errors}
q_S=-\frac{1}{2}\frac{\partial^2 F}{\partial \delta^2}\Big|_{\delta=0},
\end{equation}
we have,
\begin{equation}
\label{q1}
  q_S= \frac{1}{2}\left|\int_0^T dt (-\dot\beta\sin\theta-i\dot\theta) e^{i m (t)}\right|^2,
\end{equation}
with $m (t) =- \gamma_1$. This quantity resembles the fidelity susceptibility \cite{Gu}, which is the second order derivative of the fidelity with respect to
$\delta$, and  describes the response of the fidelity to a small error. So, to minimize or nullify the systematic error sensitivity
can somehow improve the fidelity.

In the simplest case of flat $\pi$ pulse, we have $\theta = \pi t/T$ and $\beta= -\pi/2$, therefore,
the error sensitivity gives $ q_S = \pi^2/2$, independence of $T$. It is consistent with other results that
the error sensitivity is only relevant to boundary conditions but irrelevant to the duration time $T$ \cite{2012njp,PRALu,Santos}.

Inspired by \cite{PRALu,PRLDijon}, we assume that
\begin{equation}
\label{m}
  m (t)= 2 \theta+ 2 \alpha\sin{2\theta},
\end{equation}
to nullify the error sensitivity (\ref{q1}), by choosing an appropriate $\alpha$. After substituting
Eq. (\ref{phase}) into the equation, $\gamma_1 = -2( \theta+  \alpha\sin{2\theta})$, we can solve for
\begin{equation}
\label{beta}
\beta=-\arccot [2(1+2\alpha\cos{2\theta})\sin\theta].
\end{equation}
To achieve the goal of spin flip, the form of $\theta$ is assumed to be
\begin{equation}
\label{thetaf}
  \theta =3\pi \left(\frac{t}{T}\right)^2-2\pi \left(\frac{t}{T}\right)^3,
\end{equation}
from which we obtain $\alpha=0.125$ in Eq. (\ref{beta}) to make $q_S =0$.
By using Eqs. (\ref{dotflip-1}) and (\ref{dotflip-2}), the two components of magnetic field, $\Omega$ and $\Delta$, can be further inferred
for optimal shortcut with respect to the systematic error.

\begin{figure}[t]
\begin{center}
\scalebox{0.6}[0.6]{\includegraphics{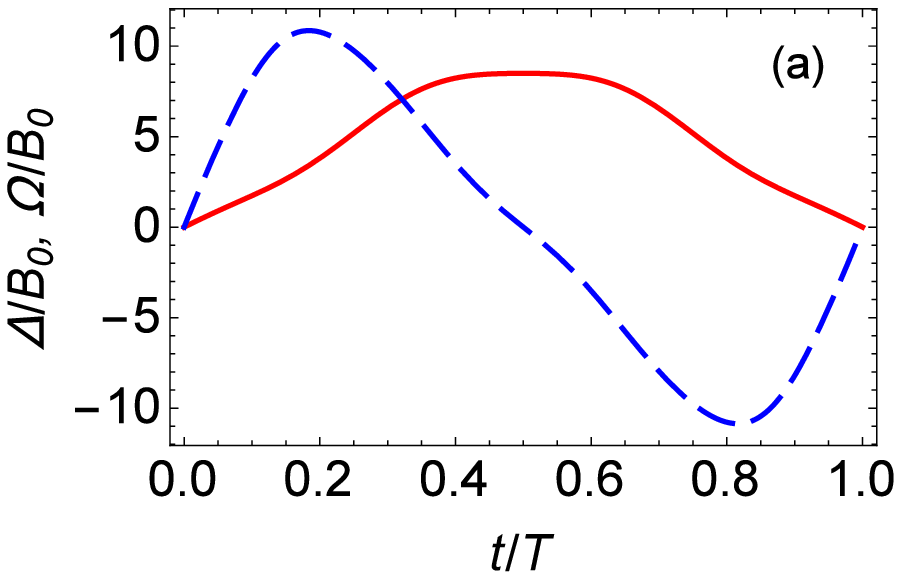}}
\\
\scalebox{0.6}[0.6]{\includegraphics{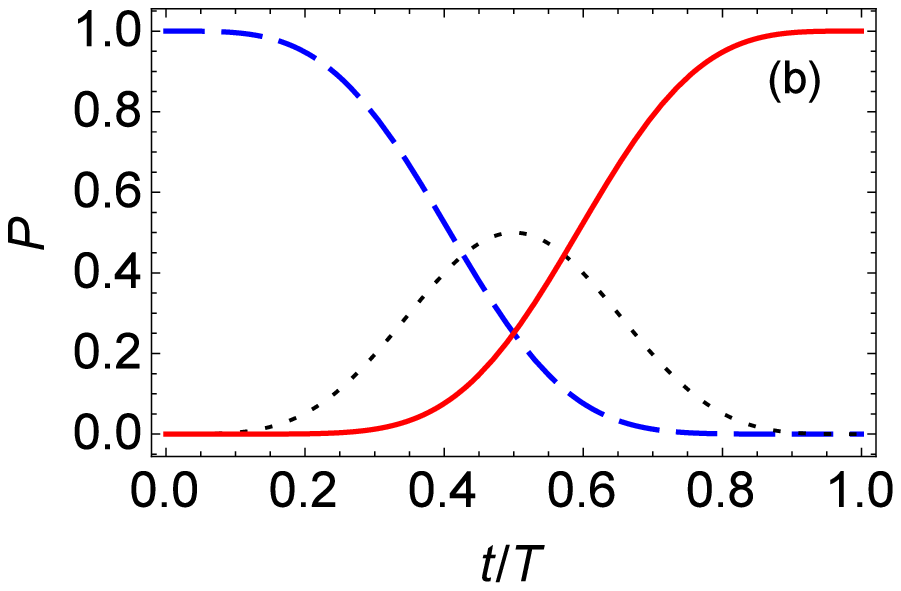}}
\\
\scalebox{0.6}[0.6]{\includegraphics{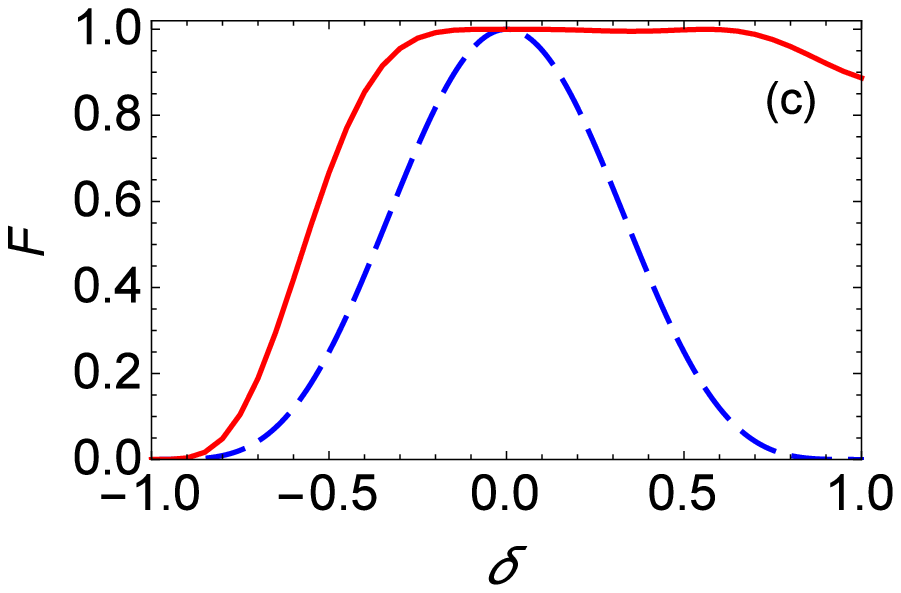}}
\caption{Optimal shortcut for spin flip in two Heisenberg-interacting spins with rotating magnetic fields, by taking into account the systematic error.
(a) Optimal protocol of two components of magnetic field, $\Omega$ (red, solid line) and $\Delta$ (blue, dashed line), in the unit of $B_0$.
(b) Dynamics of spin, described by the probability of state $|\psi_{1,1}\rangle$ (blue, dashed line), $|\psi_{1,0}\rangle$ (black, dotted line) and $|\psi_{1,-1}\rangle$ (red, solid line).
(c) Fidelity versus the systematic error $\delta$ in the amplitude of rotating magnetic field, where the optimal shortcut (red, solid line) and
flat $\pi$ pulse (blue, dashed line) are compared. Parameters: $T=1$ and $\alpha=0.125$. }\label{fig.flip}
\end{center}
\end{figure}

Figure \ref{fig.flip} presents the optimal protocol of spin flip in such two-coupled spins. The two components, $\Omega$ and $\Delta$,
of rotating magnetic field, see Fig. \ref{fig.flip} (a), are smooth enough to implement easily in practice. By using
designed magnetic field, the spin dynamics is also illustrated in Fig. \ref{fig.flip} (b), where the spin flip from $| \psi_{1,1} \rangle$ to $| \psi_{1,-1} \rangle$
is perfectly achieved. Moreover, Fig. \ref{fig.flip} (c) illustrates that the fidelity is better than flat $\pi$ pulse, where
the fidelity is calculated by solving the time-dependent Schr\"{o}dinger equation with Hamiltonian (\ref{Ha}), by using the designed magnetic fields.

\subsection{Dzyaloshinskii-Moriya interaction}

Next, we shall consider the optimal shortcuts, by taking account of DM term,
since the non-negligible magnitude results in the control errors in quantum information processing \cite{Guerrero}.
This DM interaction, originally introduced by Dzyaloshinskii \cite{Dzy} and Moriya \cite{Moriya}, is anisotropic antisymmetric exchange interaction
arising from the spin-orbit coupling, and has the following form:
\begin{equation}\label{HDM}
 H_{DM}(t) = \vec{D} \cdot(\vec{S}_1 \times \vec{S}_2),
\end{equation}
where $\vec{D}$ can be simple chosen as the constant DM vector $D$ along the $z$ axis. In the basis
of $\{S_z, S^2\}$, the Hamiltonian is written as:
\begin{equation}
  H_{DM}(t)=\left(
         \begin{array}{cccc}
           0 & 0 & 0 & 0 \\
           0 &  0 &  -\frac{i}{2}D & 0 \\
           0 & \frac{i}{2}D &0 & 0 \\
           0 & 0 & 0 & 0 \\
         \end{array}
       \right),
\end{equation}
which leads to the total Hamiltonian $\tilde{H} (t)= H(t) + H_{DM} (t)$, by combing with Eq. (\ref{Ha}),
\begin{equation}
\label{H1DM}
  \tilde{H}(t)=\left(
         \begin{array}{cccc}
          \Delta  & \frac{1}{\sqrt{2}} \Omega  & 0 & 0 \\
           \frac{1}{\sqrt{2}} \Omega  &  0 &  -\frac{i}{2}D & \frac{1}{\sqrt{2}}
           \Omega  \\
           0 & \frac{i}{2}D &  -J & 0 \\
           0 & \frac{1}{\sqrt{2}} \Omega  & 0 &  - \Delta \\
         \end{array}
       \right).
\end{equation}
Obviously, the DM interaction has the imaginary couple, which
plays  the same role as counter-diabatic driving, for example, when flipping the spin from $|\uparrow \downarrow \rangle$ to $|\downarrow \uparrow \rangle$ \cite{Shi17}. But
in this case, we would like to design the shortcut to adiabatic state evolution along the dynamical mode $|\Psi_1 (t) \rangle$
for achieving the spin flip from $|\psi_{1,1} \rangle$ to $|\psi_{1,-1} \rangle$. If one looks at the dynamical mode in Eq. (\ref{phi3-2}),
the state $| \psi_{1,0} \rangle$ is involved and populated during the shortcut path.
Therefore, such additional coupling between $| \psi_{1,0} \rangle$ and $| \psi_{0,0} \rangle$ definitely induces
an unwanted transition, and finally lowers the fidelity as a consequence.

Now, we shall apply the STA, combining with optimization, by reducing
the effect of the DM interaction. According to LR dynamical invariant theory,
the solutions of the time-dependent Schr\"{o}dinger equation with $D = 0$ are
the set of orthogonal solutions,
\begin{eqnarray}
  |\Psi_{0}(t)\rangle &=\frac{1}{\sqrt2}& \left(
                                            \begin{array}{c}
                                              -\sin{\theta}e^{-i\beta} \\
                                              \sqrt{2}\cos{\theta} \\
                                              \sin{\theta}e^{i\beta} \\
                                              0\\
                                            \end{array}
                                          \right) e^{i \gamma_0},
\end{eqnarray}
\begin{eqnarray}
  |\Psi_{1}(t)\rangle &=& \left(
                        \begin{array}{c}
                          \cos^2{\frac{\theta}{2}}e^{-i\beta} \\
                          \frac{1}{\sqrt{2}}\sin\theta \\
                          \sin^2{\frac{\theta}{2}}e^{i\beta} \\
                          0\\
                        \end{array}
                      \right)e^{i \gamma_1},
  \end{eqnarray}
\begin{eqnarray}
  |\Psi_{2}(t)\rangle &=& \left(
                        \begin{array}{c}
                          \sin^2{\frac{\theta}{2}}e^{-i\beta} \\
                         -\frac{1}{\sqrt{2}}\sin\theta \\
                          \cos^2{\frac{\theta}{2}}e^{i\beta} \\
                          0\\
                        \end{array}
                      \right)e^{i \gamma_2},
\end{eqnarray}
\begin{eqnarray}
  |\Psi_{3}(t)\rangle &=& \left(
                        \begin{array}{c}
                        0\\
                        0\\
                        0\\
                        e^{i J t}\\
                        \end{array}
                        \right),
\end{eqnarray}
where the ancillary functions of $\theta$, $\beta$ and $\gamma$ fulfil Eqs. (\ref{phase}-\ref{dotflip-2})
with the boundary conditions (\ref{bc1}) and (\ref{bc2}).

We treat the DM interaction as perturbation, and apply time-dependent perturbation theory
to obtain estimated fidelity to find the final state $|\psi_{1,-1}\rangle$ from initial state $|\psi_{1,1}\rangle$ along the state evolution $|\Psi_{1}(t)\rangle$,
\begin{equation}
\label{PDM}
 F \simeq 1-\sum_{n \neq 1} \left|\int_0^T dt\langle\Psi_1 (t)|H_{DM}(t)| \Psi_n (t)\rangle\right|^2,
\end{equation}
from which the definition of transition sensitivity \cite{Kiely14},
\beq
  q_D = -\frac{1}{2}\frac{\partial^2 F}{\partial D^2} \Big|_{D=0},
\eeq
is calculated as
\beqa
\label{qDM}
  q_D =  \frac{1}{8}\left|\int_0^T dt \sin\theta(t) e^{iJt}e^{im(t)}\right|^2,
\eeqa
with $m(t)=-\gamma_1(t)$ as before. This quantifies how sensitive a given
protocol is concerning the unwanted transition to the state $|\psi_{0,0} \rangle$.
Similar to previous part, we can have the same functions of $\theta$ and $\beta$ as Eqs. (\ref{beta}) and (\ref{thetaf}),
by imposing $m(t)$, see Eq. (\ref{m}). The numerical calculation gives $\alpha =0.059$ to nullify the transition sensitivity (\ref{qDM}), i.e. $q_D =0$.

\begin{figure}[t]
\begin{center}
\scalebox{0.6}[0.6]{\includegraphics{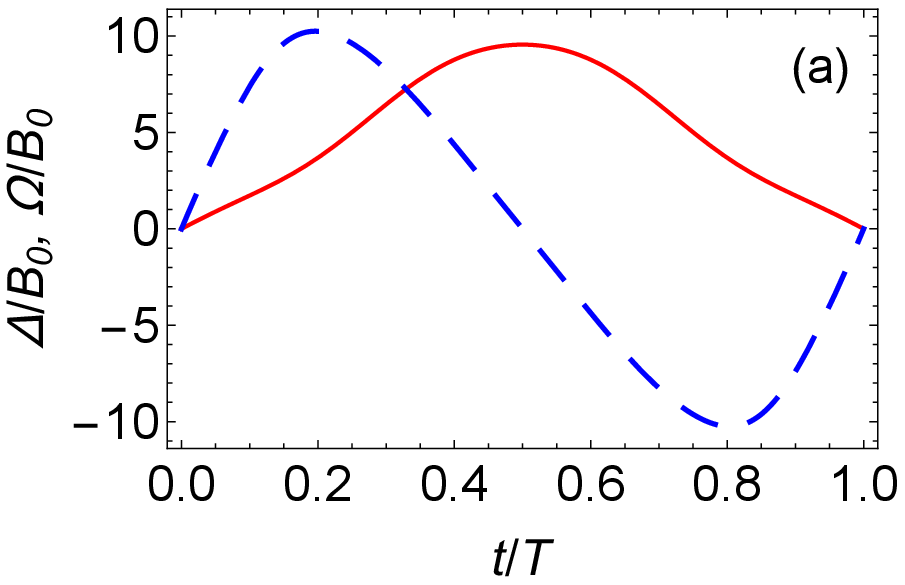}}
\\
\scalebox{0.6}[0.6]{\includegraphics{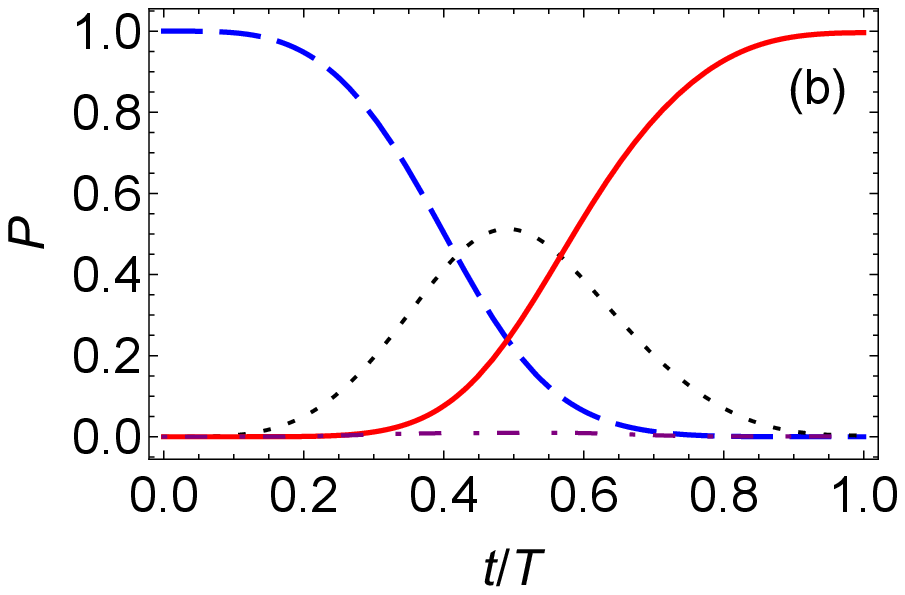}}
\\
\scalebox{0.6}[0.6]{\includegraphics{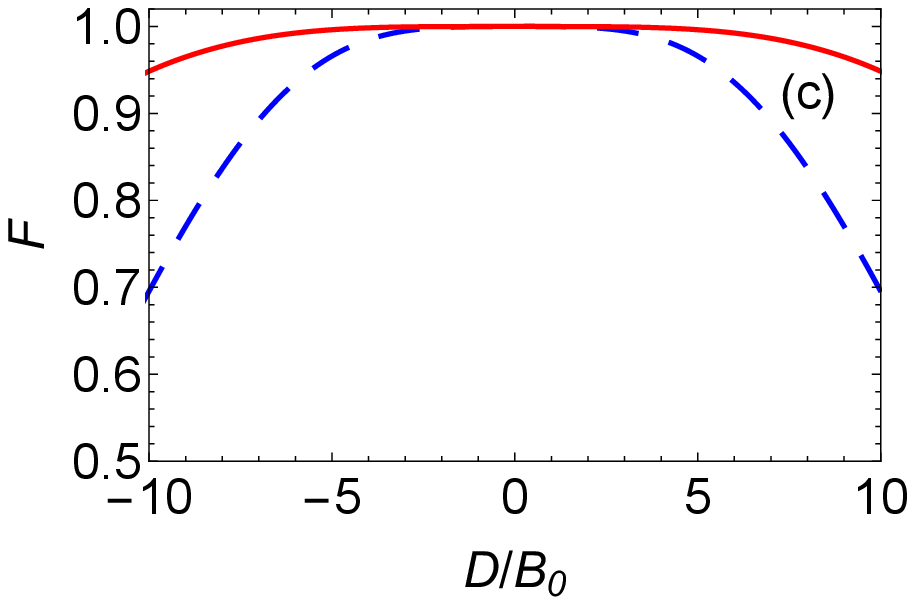}}
\caption{Optimal shortcut for spin flip in two Heisenberg-interacting spins with rotating magnetic fields,  by taking into account the perturbative DM interaction.
(a) Optimal protocol of two components of magnetic field, $\Omega$ (red, solid line) and $\Delta$ (blue, dashed line) in the unit of $B_0$.
(b) Dynamics of spin, described by the probability of state $|\psi_{1,1}\rangle$ (blue, dashed line), $|\psi_{1,0}\rangle$ ( black, dotted line), $|\psi_{0,0}\rangle$ (purple, dot-dashed line) and $|\psi_{1,-1}\rangle$ (red, solid line).
(c) Fidelity versus constant vector $D$, the magnitude of such anisotropic antisymmetric exchange, where the optimal shortcut (red, solid line) and
flat $\pi$ pulse (blue, dashed line) are compared. Parameters: $J/B_0=10$, $D/B_0=1$, $T=1$ and $\alpha=0.059$.}
  \label{fig.dmflip}
\end{center}
\end{figure}

Figure \ref{fig.dmflip} (a) and (b) demonstrates the shortcut to adiabatic spin flip driven by the optimal
magnetic fields, with two components $\Omega$ and $\Delta$, with respect to the influence of DM term.
By using designed protocol, we solve the time-dependent Schr\"{o}dinger equation with Hamiltonian (\ref{H1DM}). As a result,
the unwanted transition to the state  $|\psi_{0,0} \rangle$ can be efficiently suppressed,
and the fidelity for spin flip can be above $>0.9999$, when $D/B_0 \leq 2$. In addition,
Fig. \ref{fig.dmflip} (c) shows that the robustness of our designed protocol is better than flat $\pi$ pulse,
in which $q_D$ is constant, $q_D = 1/2\pi^2$, irrelevant to final time $T$. As a matter fact, the optimal shortcut
is quite insensitive to DM interaction, since the fidelity is reasonable when DM interaction is less than
the magnitude of magnetic fields, $\Omega/B_0 \simeq 10$.

\section{Two Ising-interacting Spins}
\label{Ising}

In this section, we shall consider the two coupled spins, described by a simple
transverse Ising model, a minimum model for quantum annealing \cite{Masuda,Takahashi-2}.
The anisotropic interaction allows us to generate the entangled Bell state, i.e. $| \psi_{1,0} \rangle$, which is crucial
in quantum information processing with atoms or spins \cite{bell01,Sarmaentangle}.
The Hamiltonian has the simple form,
\begin{equation}\label{H2}
  H (t) = J S_1^z \cdot S_2^z + \vec{B}(t)\cdot(\vec{S}_1+\vec{S}_2), 
\end{equation}
where $J$ is the exchange interaction, $\vec{S}_1$ and $\vec{S}_2$ are the two respective spin operators and  $\vec{B}(t)$ is the time-dependent rotating magnetic field. Similar to the section \ref{Heisenberg}, in the basis of $ \{S_z, \vec{S}^2 \}$, the Hamiltonian is rewritten as
\begin{equation}
\label{bellH2}
  H (t) =\left(
         \begin{array}{cccc}
           At & \frac{\Omega}{\sqrt{2}}e^{-i\omega t} & 0 & 0 \\
           \frac{\Omega}{\sqrt{2}}e^{i\omega t} & - J/2 & 0 & \frac{\Omega}{\sqrt{2}}e^{-i\omega t} \\
           0 & 0 & - J/2 & 0 \\
           0 & \frac{\Omega}{\sqrt{2}}e^{i\omega t} & 0 & -At \\
         \end{array}
       \right),
\end{equation}
in which the magnetic field components we choose are: $B_x=\Omega\cos{\omega t}$, $B_y=\Omega\sin{\omega t}$, $B_z= At$. Here the time dependence is omitted for simplicity.
Obviously, one of the state is decoupled from other three, thus the Hamiltonian, after the phase transformation, can be reduced to \cite{bell01}
\begin{equation}\label{bellH3}
  H (t)=\left(
         \begin{array}{ccc}
           -\omega+At & \frac{1}{\sqrt{2}}\Omega & 0 \\
           \frac{1}{\sqrt{2}}\Omega & -\frac{J}{2} & \frac{1}{\sqrt{2}}\Omega \\
           0 & \frac{1}{\sqrt{2}}\Omega & \omega-At \\
         \end{array}
       \right),
\end{equation}
from which we learn that level crossing happens at three different times,
$t_{12}=(\omega-\frac{J}{2})/A$, $t_{13}=\omega/A$, $t_{23}=(\omega+\frac{J}{2})/A$.
Actually, the energies of the adiabatic states have avoided crossings at certain times,
which helps us treat them as an effective two-level problem, see the details in Refs. \cite{Sarmaentangle,bell01}.

To prepare the entangled Bell state, we choose $|\psi_{1,1}\rangle$ and $|\psi_{1,0}\rangle$ as initial and final state respectively.
So, in order to avoid other crossing except for these two states, we can choose a $\Omega$ centered at $t_{12}$.
And the Hamiltonian, involving only these two states, can be eventually simplified as
\begin{equation}
\label{Hb}
  H (t)= \frac{1}{2} \left(
         \begin{array}{cc}
           \Delta & \sqrt2 \Omega \\
           \sqrt2\Omega & -\Delta \\
         \end{array}
       \right),
\end{equation}
where $\Delta=At-\omega+J/2$. Here $\Omega$ and $\Delta$ are slightly different from the previous section, which
refer to the amplitude and frequency of magnetic field.
Typically, the adiabatic passage for transferring state from $|\psi_{1,1}\rangle$ to $|\psi_{1,0}\rangle$ requires
time scale of $30/J$ \cite{QiSci,Sarmaentangle}, which gives $T^{ad} \approx 3 $ when $J=10$. Later, we shall apply STA
to shorten the time, and choose $T=1$.

For the effective two-level system, there exists dynamical invariants as follows \cite{2012njp,PRALu},
\begin{equation}\label{I}
 I(t)=\frac{B_0}{2}
\left(
                \begin{array}{cc}
                  \cos\theta & \sin\theta e^{-i\beta} \\
                  \sin\theta e^{i\beta} & -\cos\theta
                \end{array}
\right),
\end{equation}
where $\theta$ and $\beta$ are the polar and azimuthal angles in Bloch sphere. The eigenstates are
\begin{eqnarray}
\label{2solution-1}
|\phi_{+} (t)\rangle=\left(\begin{array}{c}
                      \cos\frac{\theta}{2}e^{-i\beta} \\
                      \sin\frac{\theta}{2}
                    \end{array}
                    \right),~
\label{2solution}
|\phi_{-} (t)\rangle=\left(\begin{array}{c}
                      \sin\frac{\theta}{2} \\
                     - \cos\frac{\theta}{2}e^{i\beta}
                    \end{array}
                    \right), ~~~~~
\end{eqnarray}
corresponding to the eigenvalues, $\lambda_{\pm} = \pm B_0/2$.
Similar to Sec. \ref{Heisenberg}, the solution of time-dependent Schr\"{o}dinger equation, $i \hbar \partial_t
| \Psi(t) \rangle = H | \Psi(t) \rangle$, can be written as
\begin{equation}
|\Psi (t)\rangle= \Sigma_n c_n |\phi_{n}(t)\rangle e^{i\gamma_n},
\end{equation}
where $c_n$ is time-independent constant and the LR phases $\gamma_n$ are calculated as
\begin{equation}
\label{phase-2}
\gamma_{\pm} = \pm \frac{1}{2} \int_0^t dt' \left( \dot{\beta} + \frac{\dot\theta \cot\beta}{\sin\theta}\right).
\end{equation}

To satisfy the condition for dynamical invariant, the time-dependent control parameters $\theta$ and $\beta$ are connected to
$\Omega$ and $\Delta$ by the following equations
\begin{eqnarray}
\label{dotbell-1}
  \dot\theta &=& -\sqrt2 \Omega\sin\beta, \\
\label{dotbell-2}
  \dot\beta &=& \Delta-\sqrt2 \Omega\cot\theta\cos\beta,
\end{eqnarray}
from which we can design state evolution inferred for the driving fields, fulfilling the appropriate boundary conditions.
This invariant-based inverse engineering provides the efficient way to design shortcut for generating the entangled Bell state.
To do this, we assume the state evolves along the dynamical mode $|\phi_{+} (t)\rangle$, and the same boundary conditions, Eqs. (\ref{bc1}) and (\ref{bc2}),
are also imposed. Once the functions of $\theta$ and $\beta$ satisfying the boundary conditions, the driving magnetic field
can be designed to prepare the entangled Bell state from initial state $|\psi_{1,1}\rangle$ to final state $|\psi_{1,0}\rangle$ within a short time scale \cite{Sarmaentangle,QiSci}.
However, the freedom to choose the function leaves the flexibility for further optimization, which we shall fill the gap below.

\subsection{Systematic Error}

We shall consider the optimization of STA, since there may exist the systematic errors in both Rabi frequency and detuning,
simultaneously, since they refer to the control error in the amplitude and frequency of magnetic field, $\vec{B} (t)$.
Firstly, we describe the systematic error only in Rabi frequency by the following form:
\begin{equation}
  H' (t)= \frac{1}{2} \left(
         \begin{array}{cc}
           0 &  \sqrt2\delta \Omega\\
            \sqrt2\delta \Omega & 0 \\
         \end{array}
       \right),
\end{equation}
which means the amplitude of magnetic field is changed as $\Omega \rightarrow \Omega(1+\delta)$ due to the fluctuation.
By using time-dependent perturbation theory, we have
\begin{equation}
\begin{split}
  &|\Psi(T)\rangle=|\Psi_{+}(T)\rangle-i \int_0^T dt U_0(T,t) H'|\Psi_{+}(t)\rangle\\
  &-\int_0^T dt \int_0^tdt'U_0(T,t)H'U_0(t,t')H'|\Psi_{+}(t')\rangle+...,
  \end{split}
\end{equation}
where the unperturbed time evolution operator $U_0(T,t)= |\Psi_{+} (T)\rangle \langle \Psi_{+} (t) |+ |\Psi_{-} (T)\rangle \langle \Psi_{-} (t) |$ with
$|\Psi_{\pm} (t) \rangle =e^{i \gamma_{\pm} (t)} |\phi_{\pm} (t) \rangle$.
\begin{figure}[t]
\begin{center}
\scalebox{0.6}[0.6]{\includegraphics{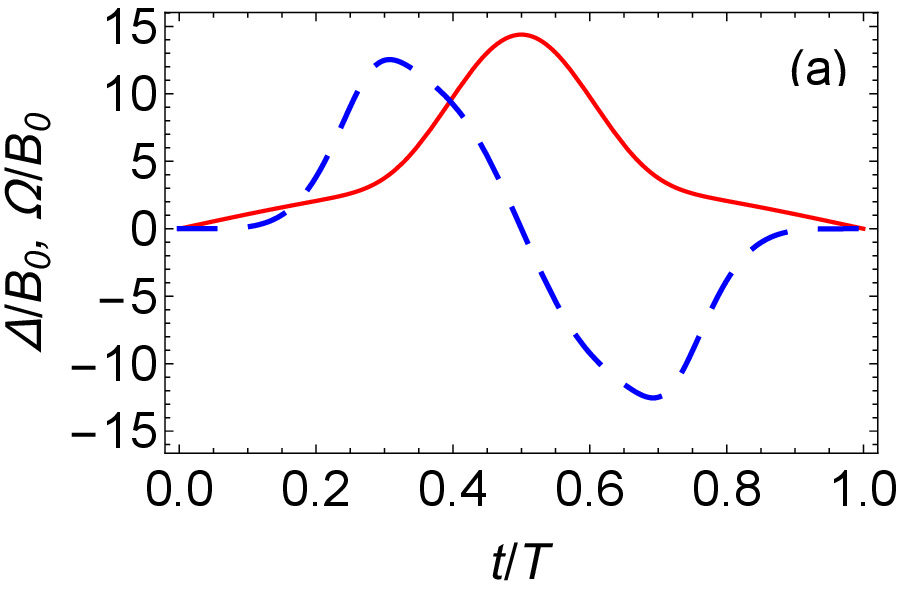}}\\
\scalebox{0.6}[0.6]{\includegraphics{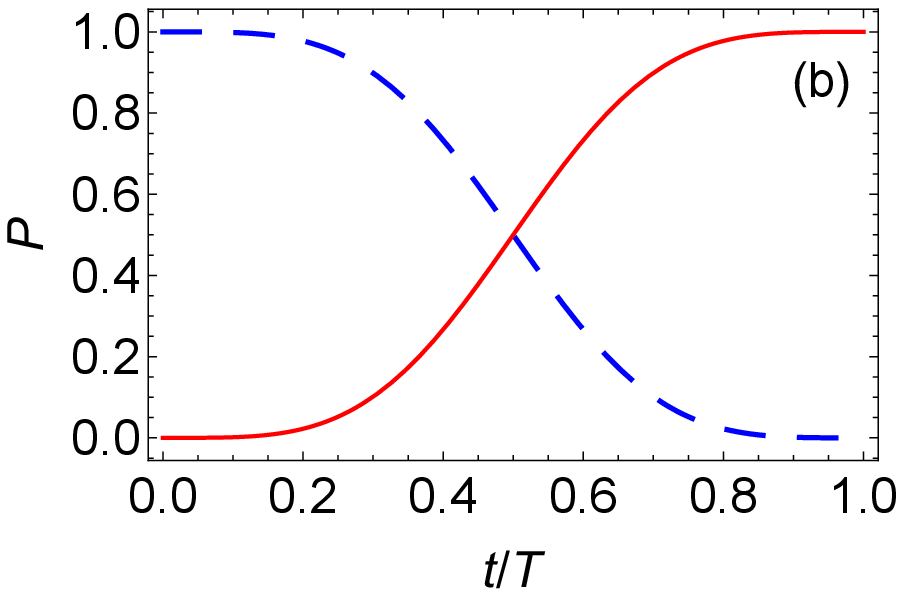}}\\
\scalebox{0.6}[0.6]{\includegraphics{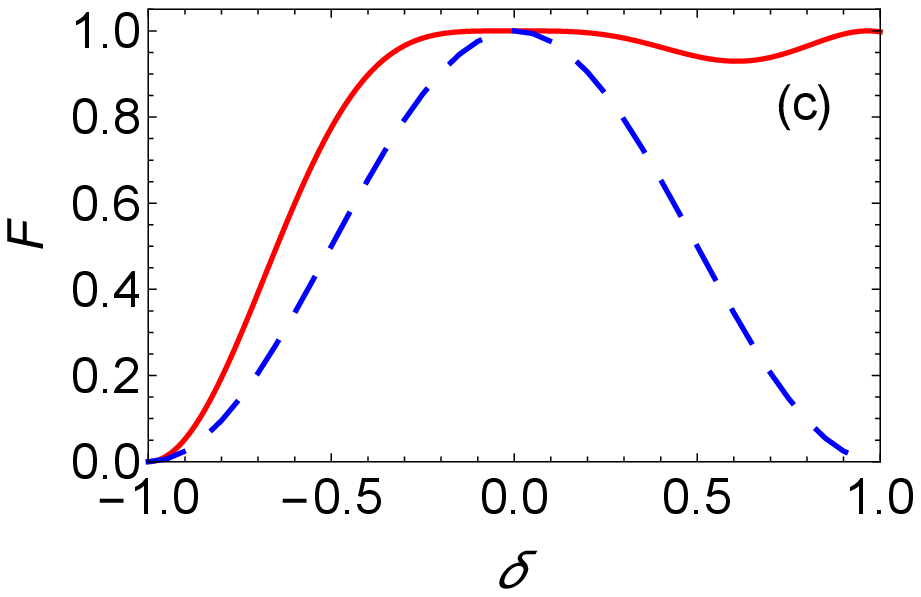}}
\caption{Optimal designed protocol for fast generation of the entangled Bell state in two Ising-interacting spins with driving magnetic fields, by taking into account
the amplitude error. (a) Optimal protocol of time-dependence $\Omega$ (red, solid line) and $\Delta$ (blue, dashed line), in the unit of $B_0$.
(b) Dynamics of spin, described by the probability of state $|\psi_{1,1}\rangle$ (blue, dashed line) and $|\psi_{1,0}\rangle$ (red, solid line).
(c) Fidelity versus the systematic error $\delta$, comparing the optimal shortcut (red, solid line) with the flat $\pi$ pulse (blue, dashed line).
Parameter: $T=1$.}
  \label{fig.bell05}
\end{center}
\end{figure}
Accordingly, we can calculate the estimated fidelity to be found at the final state $|\psi_{1,1}\rangle$ from initial state $|\psi_{1,0}\rangle$,
\beq
\label{p2}
F \simeq 1- \left|\int_0^T dt\langle\Psi_{+} (t)| H' |\Psi_{-} (t)\rangle\right|^2.
\eeq
Similarly, by defining the error sensitivity \cite{2012njp},
\begin{equation}
q_{\Omega}=-\frac{1}{2}\frac{\partial^{2} F}{\partial\delta^{2}}\bigg|_{\delta=0},
\end{equation}
we will have
\begin{equation}
\label{qnomega}
q_{\Omega} = \frac{1}{4}\left|\int_0^T dt \dot{\theta} \sin^2\theta e^{i m(t)}\right|^2,
\end{equation}
with
$
m(t)=2\gamma_{-}-\beta
$.
Thus the fidelity is approximated as $F\simeq 1- q_\Omega \delta^2$, keeping the second order.
In the special case, $m(t)=0$, $q_{\Omega} = \pi^2/4$, independently of $T$. With the choice of
$\beta = -\pi/2$, the result will recover the flat $\pi$ pulse.

Following \cite{2012njp,PRLDijon}, we can obtain $q_{\Omega} = \sin^2(n \pi)/(4 n^2)$ by imposing
$
m(t) = n (2 \theta-\sin2\theta)
$,
so that
$q_{\Omega} =0$ is achieved, if $n=1,2,3...$. In the case of $n=1$, we have
\beq
\label{beta1}
\beta= -  \arccot (4 \sin^{3} \theta),
\eeq
which results in $q_{\Omega} =0$.
When $q_{\Omega}$ is nullified,
the minimum of $q_{\Omega}$ is achieved, which gives the maximal robustness with respect to amplitude variations of magnetic field.
As an example, we choose smooth function
\beq
\label{theta}
\theta=\frac{\pi}{2}\left[1 + \sin\frac{\pi(2t-T)}{2T}\right],
\eeq
satisfying Eqs. (\ref{bc1}) and (\ref{bc2}). Notice that we use the different ansatz from previous one, to show the versatility.
As a consequence, we can solve for $\Omega$ and $\Delta$ through Eqs. (\ref{dotbell-1}) and (\ref{dotbell-2}).

Figure \ref{fig.bell05} illustrates the optimal design of driving magnetic field with respect to amplitude error, and corresponding spin evolution
from $|\psi_{1,1} \rangle$ to $|\psi_{1,0} \rangle$. The dynamics and fidelity are calculated numerically
by solving the time-dependent Schr\"{o}dinger equation with Hamiltonian (\ref{Hb}).
The comparison between the shortcut protocol and flat $\pi$ pulse is made to show the maximum robustness.
Besides, we can also optimize the shortcut with respect to the systematic error in detuning, as in Ref. \cite{PRALu}. We will discuss the similar situation below.

\subsection{DM Interaction}

We turn to consider the effect DM interaction on the Bell state generation and further reduce the control error by optimizing STA.
After phase transformation, the total Hamiltonian, with the DM term, is written as
\begin{equation}
  \tilde{H} (t) =\left(
         \begin{array}{cccc}
           B_z & \frac{\Omega}{\sqrt{2}} e^{- i \omega t} & 0 & 0 \\
           \frac{\Omega}{\sqrt{2}}  e^{i \omega t}   & -J/2 & -iD & \frac{\Omega}{\sqrt{2}}  e^{- i \omega t}   \\
           0 & iD &  -J/2 & 0 \\
           0 & \frac{\Omega}{\sqrt{2}}  e^{i \omega t}  &  0 & -B_z \\
         \end{array}
       \right),
\end{equation}
from which we can further obtain the effective Hamiltonian for three-level system, by adiabatic elimination \cite{Li},
\begin{equation}
\label{HDM3}
  \tilde{H} (t) =\left(
         \begin{array}{ccc}
           B_z & \frac{\Omega}{\sqrt{2}} e^{- i \omega t}  & 0 \\
           \frac{\Omega}{\sqrt{2}}  e^{i \omega t}   & -J/2+\delta & \frac{\Omega}{\sqrt{2}}  e^{- i \omega t}   \\
           0 & \frac{\Omega}{\sqrt{2}}  e^{i \omega t}  & -B_z \\
         \end{array}
       \right),
\end{equation}
with $\delta = 2 D^2/J$.
By choosing $B_x=\Omega\cos{\omega t}$, $B_y=\Omega\sin{\omega t}$, $B_z= At$ as before, the Hamiltonian, after the phase transformation, becomes
\begin{equation}
\label{bellH3DM}
  \tilde{H} (t)=\left(
         \begin{array}{ccc}
           -\omega+At & \frac{1}{\sqrt{2}}\Omega & 0 \\
           \frac{1}{\sqrt{2}}\Omega & -\frac{J}{2}+\delta & \frac{1}{\sqrt{2}}\Omega \\
           0 & \frac{1}{\sqrt{2}}\Omega & \omega-At \\
         \end{array}
       \right),
\end{equation}
from which we learn that level crossing happens at three different times,
$t_{12}=(\omega-\frac{J}{2}+\delta)/A$, $t_{13}=\omega/A$, $t_{23}=(\omega+\frac{J}{2}-\delta)/A$.
By comparing the Hamiltonian (\ref{bellH3DM}) with (\ref{bellH3}), we finally find
the Hamiltonian, involving only two states of $|\psi_{1,1}\rangle$ and $|\psi_{1,0}\rangle$, as follows
\begin{equation}
  \tilde{H} (t)= \frac{1}{2} \left(
         \begin{array}{cc}
           \Delta -\delta & \sqrt2 \Omega \\
           \sqrt2\Omega & -\Delta + \delta \\
         \end{array}
       \right),
\end{equation}
where $\Delta$ is the same as before. As a result, the DM term in this case can be considered as the shift of detuning,
and described by the perturbative Hamiltonian
\beq
  H'(t)= \frac{1}{2} \left(
         \begin{array}{cc}
           -\delta & 0 \\
           0 & \delta \\
         \end{array}
       \right).
\eeq

\begin{figure}[t]
\begin{center}
\scalebox{0.6}[0.6]{\includegraphics{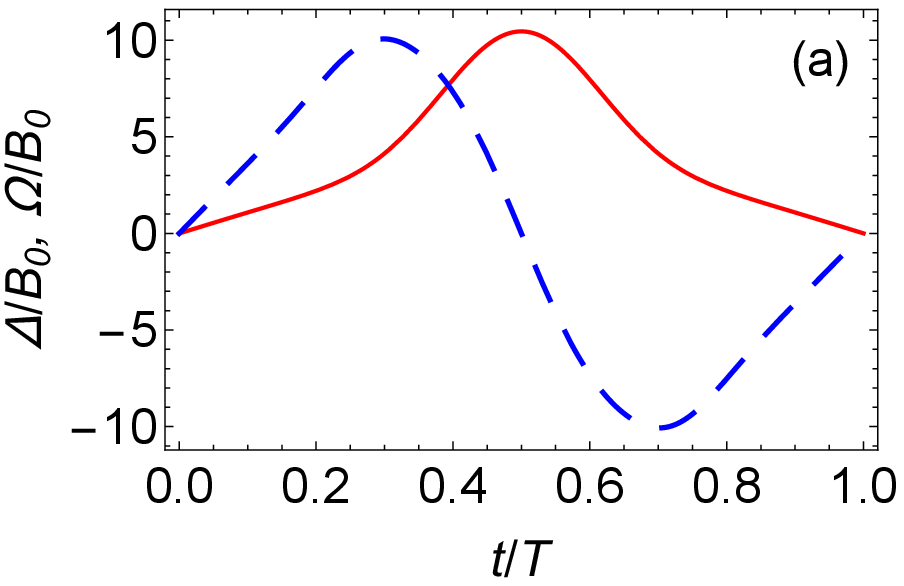}}\\
\scalebox{0.6}[0.6]{\includegraphics{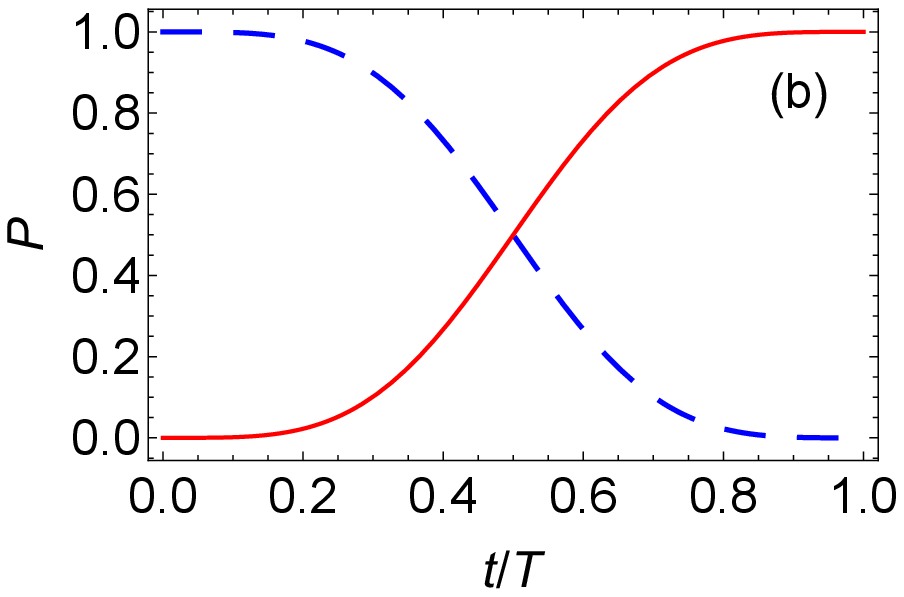}}\\
\scalebox{0.6}[0.6]{\includegraphics{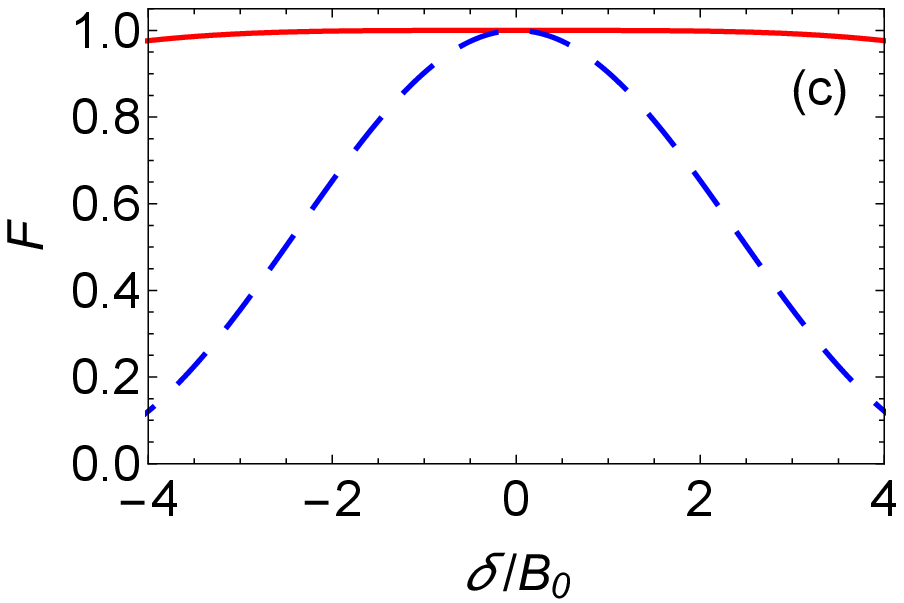}}
\caption{
Optimal designed protocol for fast generation of the entangled Bell state in two Ising-interacting spins with driving magnetic fields, by taking into account
the DM interaction. (a) Optimal protocol of time-dependence $\Omega$ (red, solid line) and $\Delta$ (blue, dashed line), in the unit of $B_0$.
(b) Dynamics of spin, described by the probability of state $|\psi_{1,1}\rangle$ (blue, dashed line) and $|\psi_{1,0}\rangle$ (red, solid line).
(c) Fidelity versus the systematic error $\delta$, relevant to DM term, comparing the optimal shortcut (red, solid line) with the flat $\pi$ pulse (blue, dashed line).
Parameters: $J/B_0=10$, $D/B_0=1$, $T=1$ and $\alpha=-0.206$.}
  \label{fig.bellDM}
\end{center}
\end{figure}

Similar to Eqs. (\ref{p2}) and (\ref{qnomega}), we can finally obtain \cite{PRALu}
\begin{equation}
\label{qndelta}
q_{\Delta} =\frac{1}{4}\left|\int_0^T dt \sin\theta e^{im(t)}\right|^2.
\end{equation}
Again, we may assume $ m(t)$ in Eq. (\ref{m}) to nullify $q_{\Delta}$, with free parameter $\alpha$.
After some straightforward calculations, we finally obtain $q_{\Delta}=0$ with
the parameter $\alpha=-0.206$. Fig. \ref{fig.bellDM} demonstrates the optimal protocol and corresponding spin dynamics
designed by STA, in which the flat $\pi$ pulse, $q_{\Delta} = (T/\pi)^2$, is also compared.
Since $\delta= 2D^2/J$, the error induced by DM could be negative and positive depending on
antiferromagnetic coupling $J>0$ or ferromagnetic coupling $J<0$. When $\delta=4$, we have $D =14.4$ in the unit of $B_0$, with the parameter $J=10$.
Thus, it is proved, by solving the time-dependent Schr\"{o}dinger equation with Hamiltonian (\ref{Hb}),
that our designed protocol is insensitive to the DM interaction, see Fig. \ref{fig.bellDM} (c), when $D$ is less than the magnitude of driving magnetic field.
This is extremely useful to generate the entangled Bell state in two coupled spin systems.

\section{Discussion}
\label{discussion}

\begin{figure}[t]
\begin{center}
\scalebox{0.6}[0.6]{\includegraphics{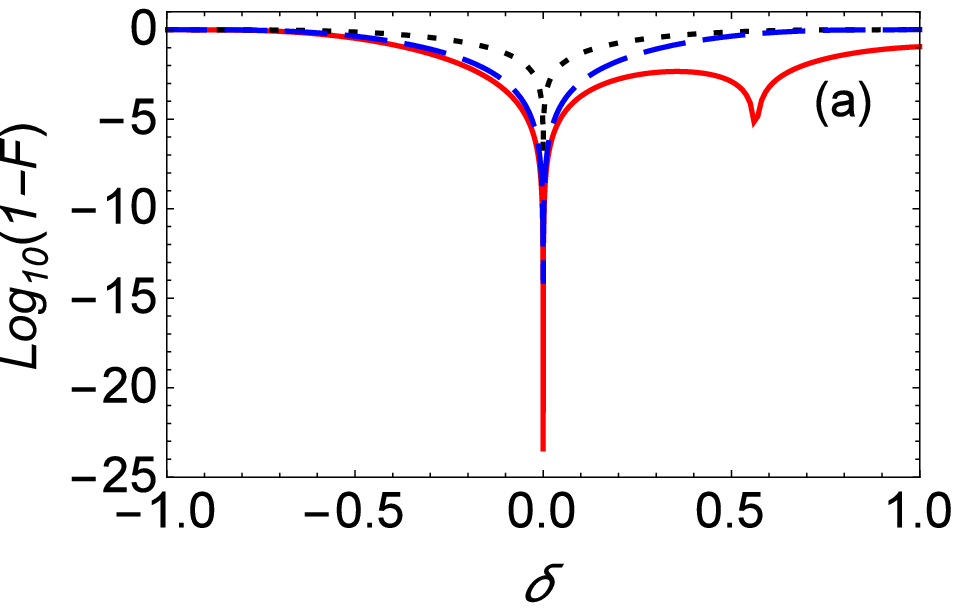}}\\
\scalebox{0.6}[0.6]{\includegraphics{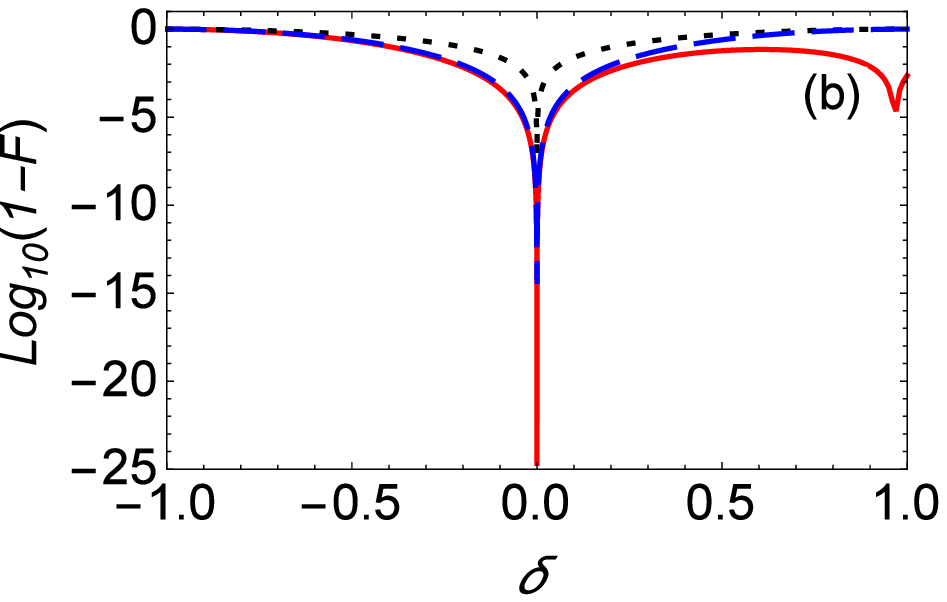}}
\caption{Robustness of designed shortcut protocols (red, solid line), by comparing with flat $\pi$ pulse (black, dotted line) and composite pulse (blue, dashed line),
where (a) Heisenberg-interacting and (b) Ising interaction are considered. Parameters: $T=1$ and others are the same as those in Figs. \ref{fig.flip} and \ref{fig.bell05}. }
\label{fig.composite}
\end{center}
\end{figure}

In this section, we shall discuss the robustness of designed shortcut protocols, by comparing with the flat $\pi$ pulse and a composite pulse $\frac{\pi}{2}(x) \pi(y)\frac{\pi}{2} (x)$. Especially, the techniques of composite pulses, originally proposed in NMR \cite{NMR} , are popular in robust high-fidelity quantum control \cite{Torosov,VitanovPRL}.
Firstly, we consider the spin-flip case when the Heisenberg interaction is present, in Sec. \ref{Heisenberg}. When the initial state
is $|\psi_{1,1}\rangle$, the state evolution can be represented as $|\psi(t)\rangle=U (t,0) |\psi_{1,1}\rangle$,
where $U(t,0)$ is the propagator, connecting the initial state with final state.
For instance, when the Hamiltonian $H = \Omega J_{i}$ ($i=x, y$), the propagator, $U_i (T,0)= \exp(- i H t)$, is calculated as
\begin{equation}
\label{US}
 U_i(t,0) =I-i\sin(\Omega t) J_i+[\cos(\Omega t)-1] J_i^2,
\end{equation}
where $J_{i}$ ($i=x,y$) are generator matrices of spin-1, as defined before.
Considering the flat $\pi$ pulse with the systematic error $\delta$, the Hamiltonian is described by $H= \Omega(1+\delta) J_x$ with $\Omega T = \pi$.
By using the propagator in Eq. (\ref{US}),
the fidelity, $F = | \langle \psi_{1,-1} |U(T,0)|  \psi_{1,1} \rangle|^2$, to be found in the desired sate $|\psi_{1,-1}\rangle$, is thus
analytically solved as,
\beq
\label{f-pi}
F = \cos^4 \left(\frac{\pi\delta}{2}\right) \simeq 1 - \frac{\pi^2\delta^2}{2}.~~~~~~~~~~~
\eeq
This recovers that $q_S = \pi^2/2$, as discussed in Sec. \ref{Heisenberg}.
For a composite pulse, we consider the typical one, $\frac{\pi}{2}(x) \pi(y)\frac{\pi}{2} (x)$,
in order to minimize resonance offset effects, where the combination $x$, $y$
refer to the two components of magnetic field and spin-1 operators, $J_x$ and $J_y$, involved \cite{NMR}.
Similarly, we can calculate the propagator by using Eq. (\ref{US}), and finally obtain the fidelity,
\beqa
\label{Fcom}
F = \cos^8\left(\frac{\pi\delta}{2}\right)+\sin^2(\pi\delta)\cos^2\left(\frac{\pi\delta}{2}\right)\simeq 1-\frac{\pi^4\delta^4}{8}.
\eeqa
Interestingly, we notice that for composite pulse the error sensitivity, defined in Eq. (\ref{q1}), are null. This suggests
that the composite pulse can improve the robustness, but more sequences require more time, as compared to the optimally robust shortcut protocol with single-shot pulse.
Fig. \ref{fig.composite} (a) demonstrates that the optimal shortcut with respect to the systematic error, as shown in Fig. \ref{fig.flip},
more robust, as compared with the flat $\pi$ pulse and composite pulse.

Regarding the generation of entangled Bell state in two Ising-interaction spins,
we consider the effective two-level Hamiltonian. In this case,
we calculate the fidelities
\begin{equation}
\label{InFpi}
  F =\cos^2 \left(\frac{\pi\delta}{2}\right) \simeq 1 - \frac{\pi^2\delta^2}{4},
\end{equation}
for flat $\pi$ pulse with $\sqrt{2} \Omega T =\pi$ and
\beq
 F =1- \sin^4 \left(\frac{\pi \delta}{2}\right) \simeq 1-\frac{\pi^4\delta^4}{16},
\eeq
for composite pulse $\frac{\pi}{2}(x) \pi(y)\frac{\pi}{2} (x)$, with Pauli matrix, $\sigma_x$ and $\sigma_y$, involved instead.
For comparison,  we calculate that $q_\Omega =\pi^2/4$ for flat $\pi$ pulse and $q_\Omega = 0$ for composite pulse, based on Eq. (\ref{qnomega}),
which are consistent with the discussion in Sec. \ref{Ising}.
Fig. \ref{fig.composite} (b) confirms that the optimally shortcut protocol, as shown in Fig. \ref{fig.bell05}, is ultrarobust against the systematic error,
and its stability is better than the flat $\pi$ pulse and composite pulse.

In this closing section, we shall emphasize the extension of the present optimal shortcuts to many-spin system.
For instance, we can consider the Hamiltonian,
\begin{equation}
\label{H3spin}
  H (t) = J \sum^{3}_{i=1}\sum_{j\neq i}S_i^z S_j^z + \vec{B}(t)\cdot (\sum^{3}_{i=1}\vec{S}_i), 
\end{equation}
describing three spins at the vertex of an equilateral triangle. By imposing the three components of magnetic field,
$B_x=\Omega\cos{\omega t}$, $B_y=\Omega\sin{\omega t}$ and $B_z=At$,
we can use the similar approach in Refs. \cite{bell01,Sarmaentangle}, and eventually write down the effective two-level Hamiltonian as follows
\begin{equation}
  H (t)= \frac{1}{2} \left(
         \begin{array}{cc}
           \Delta & \sqrt{3} \Omega \\
           \sqrt{3}\Omega & -\Delta \\
         \end{array}
       \right),
\end{equation}
with $\Delta = At + \omega + J$, in the basis of
\beqa
\label{basis-2}
  &|&\psi_{3/2,3/2}\rangle = |\uparrow\uparrow \uparrow \rangle, \\
  &|&\psi_{3/2,1/2}\rangle=
  \frac{1}{\sqrt3}(|\uparrow\uparrow\downarrow\rangle+|\uparrow\downarrow \uparrow\rangle+|\downarrow \uparrow\uparrow \rangle ).
\eeqa
This allows us to construct the dynamical invariant as before, and design inversely the components of magnetic field to achieve fast and robust
generation of W entangled state $| \psi_{3/2,1/2}\rangle$ from the initial state $| \psi_{3/2,3/2}\rangle $.
More interestingly, the same approach can be further applied to manipulate the larger number of coupled spins, when the Rabi frequency is replaced by
$\sqrt{n} \Omega$, see also Ref. \cite{QiSci}.

Besides, one can generalize the shortcut protocol to Heisenberg interaction with many spins,
of which the Hamiltonian is
\beq
H (t)=  J \sum_{(i,j} \vec{S}_i \cdot \vec{S}_j + \vec{B}(t)\cdot (\sum^{N}_{i=1}\vec{S}_i),
\eeq
where $N$ is the total spin number and $(i,j)$ stands for the summation process that we
only take account of the nearest neighbour spins. The dynamical invariant in this situation
is constructed in Ref. \cite{LiPLA},  and STA technique will be utilized as well. But if the Heisenberg interaction
is anisotropic \cite{Suter-review}, the dynamical invariant does not exist. Therefore, the above semi-analytical analysis is not applicable, and
more complicated numerical analysis is required.

\section{Conclusion}
\label{conclusion}

In summary, we have studied the fast and robust control of spin states in two interacting spins with Heisenberg and Ising interaction
by using STA technique. We first apply inverse engineering to design the time-dependent magnetic fields for spin flip, and
optimize the spin dynamics with respect to control error and fluctuation. In particular, the optimally shortcut protocol provides
an efficient way to suppress the unwanted transition or systematic error induced by DM, due to anisotropic antisymmetric exchange.
Furthermore, the anisotropic Ising interaction enables us to generate the entangled Bell state, and the fast and stable
process is interesting for quantum information processing.

We emphasize that the optimal shortcuts designed here are different from counter-diabatic driving \cite{Sarmaentangle,Shi17}
and fast-forward scaling \cite{Masuda17}, since the inverse engineering based on LR invariant provides the explicit dynamics of spins and
the flexibility for further optimization. More robust protocol can be further designed by minimizing or nullifying the high order of the approximate fidelity, calculated from
time-dependent perturbation theory. In addition, the smooth single-shot pulses are suitable
for the applications, and could be connected to the composite pulse-sequence technique with a time-dependent phase \cite{PRLDijon}.
Here we consider the special cases of the isotropic Heisenberg XXX and simple transverse Ising models
with the semi-analytical solutions, due to the symmetry. However,
a more general anisotropic Heisenberg XYZ model requires the numerical analysis and optimization \cite{Suter-review},
since the inverse engineering based on dynamical invariant does not work.
Therefore, open questions left for future work include comparing the present protocol with other methods, such as
universal broadband composite pulse sequences \cite{VitanovPRL}, incorporating the recipe of optimal control \cite{NJPSherson},
or extending to the anisotropic interacting many-spin systems for quantum annealing \cite{Takahashi-2}.
Last but not least, the results can be readily transposed to other systems, i.e. NMR, quantum dots, superconducting circuits and optomechanical systems ect.

\section*{Acknowledgment}
Q. Z. and X. C. appreciate the fruitful discussions with D. Gu\'{e}ry-Odelin.
This work is partially supported by the NSFC (11474193, 61404079), the Shuguang (14SG35),
the program of Shanghai Municipal Science and Technology Commission (18010500400 and 18ZR1415500),
and the Program for Professor of Special Appointment (Eastern Scholar).
Y.B. also acknowledges Juan de la Cierva program.

\section*{Appendix}
\label{appendix}
In this Appendix, we would like to show how to derive the effective four-level and reduced Hamiltonians for two interacting spins.
Generally speaking, the Hamiltonian of a two coupled-spin system with anisotropic Heisenberg interaction can be written as
\begin{equation}
\label{Hg}
  H(t) = \sum_i J_i S_{1}^i S_{2}^i  + \vec{B}(t)\cdot (\vec{S}_1 + \vec{S}_2),
\end{equation}
in which $J_{i}$ ($i=x,y,z$) are the coupling coefficients, and $\vec{B} (t)$ is rotating magnetic field with
three components, $B_{i}$.
In the basis of $|\uparrow\uparrow\rangle$, $|\uparrow\downarrow\rangle$,
$|\downarrow\uparrow\rangle$, and $|\downarrow\downarrow\rangle$,
the Hamiltonian can be rewritten as
\begin{equation}
  H = \left(
        \begin{array}{cccc}
          J_z/4 + B_z & \frac{B_x - i B_y}{2} & \frac{B_x - i B_y}{2} & \frac{J_x-J_y}{4} \\
          \frac{B_x + i B_y}{2} & -J_z/4 & \frac{J_x+J_y}{4} & \frac{B_x - i B_y}{2} \\
          \frac{B_x + i B_y}{2} & \frac{J_x+J_y}{4} &  -J_z/4 & \frac{B_x - i B_y}{2} \\
          \frac{J_x-J_y}{4} & \frac{B_x + i B_y}{2} & \frac{B_x + i B_y}{2} & J_z/4- B_z \\
        \end{array}
      \right).
\end{equation}
To obtain the Hamiltonian, which can be reduced to effective three-level system in Eq. (\ref{Ha}), we represent the above Hamiltonian in
the basis of singlet and triplet states, described by Eqs. (\ref{basis}-\ref{basis4}). Therefore, we have the following expression:
\begin{equation}
  H(t)=\left(
         \begin{array}{cccc}
           \frac{J_z}{4}+ B_z  & \frac{B_x - i B_y}{\sqrt{2}} & 0 & 0 \\
           \frac{B_x + i B_y}{\sqrt{2}} &  \frac{J_x +J_y-J_z}{4} & 0 & \frac{B_x - i B_y }{\sqrt{2}} \\
           0 & 0 &  -\frac{J_x +J_y+J_z}{4} & 0 \\
           0 & \frac{B_x + i B_y}{\sqrt{2}} & 0 &  \frac{J_z}{4} - B_z \\
         \end{array}
       \right).
\end{equation}
Since the state $|\psi_{0,0}\rangle$ is decoupled to others, the population does not change, up to the phase fact. Finally,
we can write down the three-level system with Hamiltonian
\begin{equation}
\label{He3}
  H(t) = \left(
           \begin{array}{ccc}
             \frac{J_z}{4}+ B_z & \frac{B_x - i B_y}{\sqrt{2}} & 0 \\
             \frac{B_x + i B_y}{\sqrt{2}} & \frac{J_x +J_y-J_z}{4} & \frac{B_x - i B_y }{\sqrt{2}} \\
             0 & \frac{B_x + i B_y}{\sqrt{2}} & \frac{J_z}{4} - B_z \\
           \end{array}
         \right).
\end{equation}
Obviously, we eventually obtain the simple Hamiltonian in Eq. (\ref{Ha}), satisfying SU(2) Lie algebra,
with corresponding dynamical invariant, just by shifting the energy $J/4$, and setting $J_x = J_y = J_z$.
Of course, we can perform the similar calculation for Ising-interaction case, as well.

\end{document}